\documentclass[preprint,3p]{elsarticle}

\usepackage[utf8]{inputenc}
\usepackage{graphicx,subfigure}
\usepackage{textcomp}
\usepackage{gensymb}
\usepackage{tabularx}
\usepackage{tabu}

\usepackage{lineno}

\author[a]{M. Mironova}
\author[a]{K. Metodiev}
\author[b]{P. Allport}
\author[c,d]{I. Berdalovic}
\author[a]{D. Bortoletto}
\author[e]{C. Buttar}
\author[c,f]{R. Cardella}
\author[c]{V. Dao} 
\author[c]{M. Dyndal} 
\author[b]{P. Freeman}
\author[c,e]{L. Flores Sanz de Acedo} 
\author[b]{L. Gonella}
\author[c]{T. Kugathasan}
\author[c]{H. Pernegger} 
\author[c]{F. Piro}
\author[a]{R. Plackett}
\author[c]{P. Riedler} 
\author[a,c]{A. Sharma}
\author[a]{E.J. Schioppa}
\author[a]{I. Shipsey}
\author[c]{C. Solans Sanchez}
\author[c]{W. Snoeys} 
\author[b]{H. Wennlöf} 
\author[a]{D. Weatherill}
\author[a]{D. Wood}
\author[b]{S. Worm} 

\address[a]{University of Oxford (UK)}
\address[b]{University of Birmingham (UK)}
\address[c]{CERN (CH)}
\address[d]{University of Zagreb (HR)}
\address[e]{University of Glasgow (UK)}
\address[f]{University of Oslo (NO)}

\title{Measurement of the relative response of TowerJazz Mini-MALTA CMOS prototypes at Diamond Light Source}

\begin{document}

\begin{abstract}
    This paper outlines the results of investigations into the effects of radiation damage in the mini-MALTA prototype. Measurements were carried out at Diamond Light Source using a micro-focus X-ray beam, which scanned across the surface of the device in 2 $\mathrm{\mu m}$ steps. This allowed the in-pixel photon response to be measured directly with high statistics. Three pixel design variations were considered: one with the standard continuous $\mathrm{n^-}$ layer layout and front-end, and extra deep p-well and $\mathrm{n^-}$ gap designs with a modified front-end. Five chips were measured: one unirradiated, one neutron irradiated, and three proton irradiated.
\end{abstract}



\date{September 2019}

\maketitle
\section{Introduction}

\begin{figure}
\centering
  \centering
  \includegraphics[width=1\textwidth]{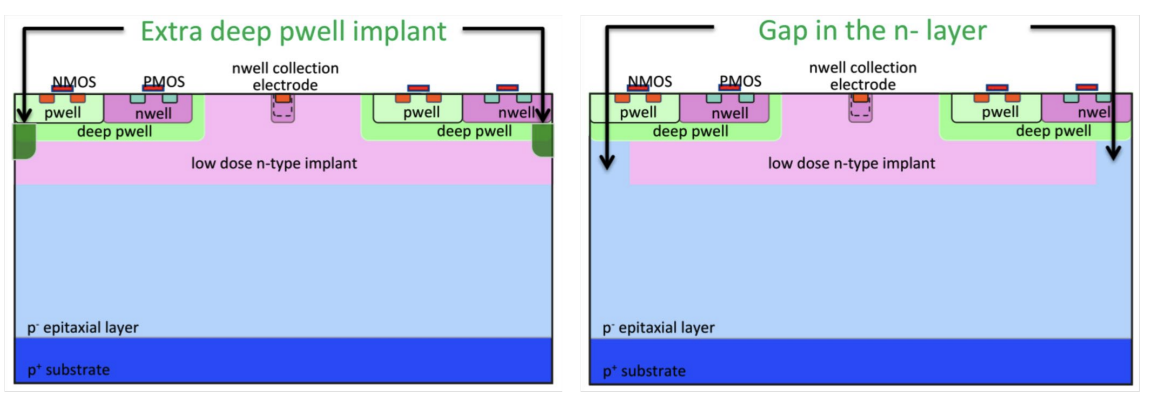}
  \caption{Process modification in mini-MALTA. Figure from \cite{miniMALTA}.}
  \label{fig:processes}
\end{figure}

The mini-MALTA device is a depleted monolithic pixel sensor prototype made in TowerJazz 180~nm CIS technology, with 64x16 pixels and 36.4 x 36.4 $\mathrm{\mu m^2}$ pixel pitch. The device, laboratory and particle testbeam measurements are described in \cite{MiniMALTA2019}. It is based on the design of the MALTA chip, which was a full-sized CMOS demonstrator \cite{TowerJazz2, TowerJazzATLAS}. Mini-MALTA has eight sectors which differ in their front-end design, reset mechanism and process. The standard design, similar to the one of the MALTA chip, is referred to as the continuous $\mathrm{n^-}$ layer design. Additionally, there are two process modifications which have been introduced in the mini-MALTA design and are shown in Figure \ref{fig:processes}. The aim of these two modifications in the substrate is to improve charge collection at the pixel edges. The first variant has an extra deep implant of p-type silicon, which should shape the electric field such that the produced charge carriers are steered more directly towards the collection electrode in the centre of the pixel. The second design has a gap in the $\mathrm{n^-}$ layer, which is expected to have a similar effect on the electric field. Both of the designs have been shown to perform well in TCAD simulations \cite{TCAD}. Each of the designs is implemented with a standard front-end and with a modified front-end featuring enlarged transistors to reduce random telegraph signal (RTS) noise. For the continuous $\mathrm{n^-}$ layer design the sector with the standard front-end was tested, and for the $\mathrm{n^-} $ gap and extra deep p-well designs, the sectors with the enlarged transistors were measured.

The main goal of the testbeam at Diamond Light Source was to study the performance of the two new designs in comparison to the standard design for different levels of irradiation. One unirradiated chip was tested (W2R11). One sample (W2R1) was irradiated at Ljubliana with 1 MeV neutrons up to 1e15 $\mathrm{n_{eq}/cm^2}$. The remaining three chips were irradiated with 23 MeV protons at the MC40 cyclotron at Birmingham to 5e14 and 7e13 $\mathrm{n_{eq}/cm^2}$ \cite{Birmingham}. Here the delivered fluence was estimated through nickel dosimetry and the total ionising dose (TID) was calculated using the Bethe model for 23 MeV protons. Table \ref{table:irradiations} summarises the properties and irradiation doses of the different chips. The devices were tested three weeks after irradiation with noise and threshold scans, as well as data acquisition with a Sr-90 source. One of the samples (W5R9) was annealed at low temperatures to obtain reasonable results for these tests, as its noise was too high after irradiation.

\begin{table}
\centering
\begin{tabularx}{\textwidth}{|X|X|X|X|X|X|}
 \hline
  Sample & Epitaxial Thickness $(\mu\mathrm{m})$ & Irradiation & Fluence  $(\mathrm{n_{eq}/cm}^2)$ & TID (MRad)  & annealing  \\ [0.5ex] 
 \hline\hline
  W2R11 & 30 &   none         &      0 & 0  & none    \\ [1ex] \hline
  W2R9  & 30 & 23 MeV protons & 5.0e14 & 66 & none    \\ [1ex] \hline
  W2R1  & 30 & 1 MeV neutrons &   1e15 &    & none    \\ [1ex] \hline
  W5R9  & 25 & 23 MeV protons & 5.6e14 & 74 & 120 min at 35 $^\circ$C  \\ [1ex] \hline
  W4R9  & 25 & 23 MeV protons &   7e13 &  9 & none    \\ [1ex] \hline
 \hline
\end{tabularx}
\caption{Table of mini-MALTA chips with their respective delivered dose in neutron equivalent fluence and TID.}
\label{table:irradiations}
\end{table}


\section{Measurements and techniques}

\begin{figure}
\centering
\subfigure[Experimental Setup in the beamline. The mini-MALTA chip is cooled and located on a PCB, which is mounted onto a motion stage.]{\includegraphics[width=0.55\textwidth]{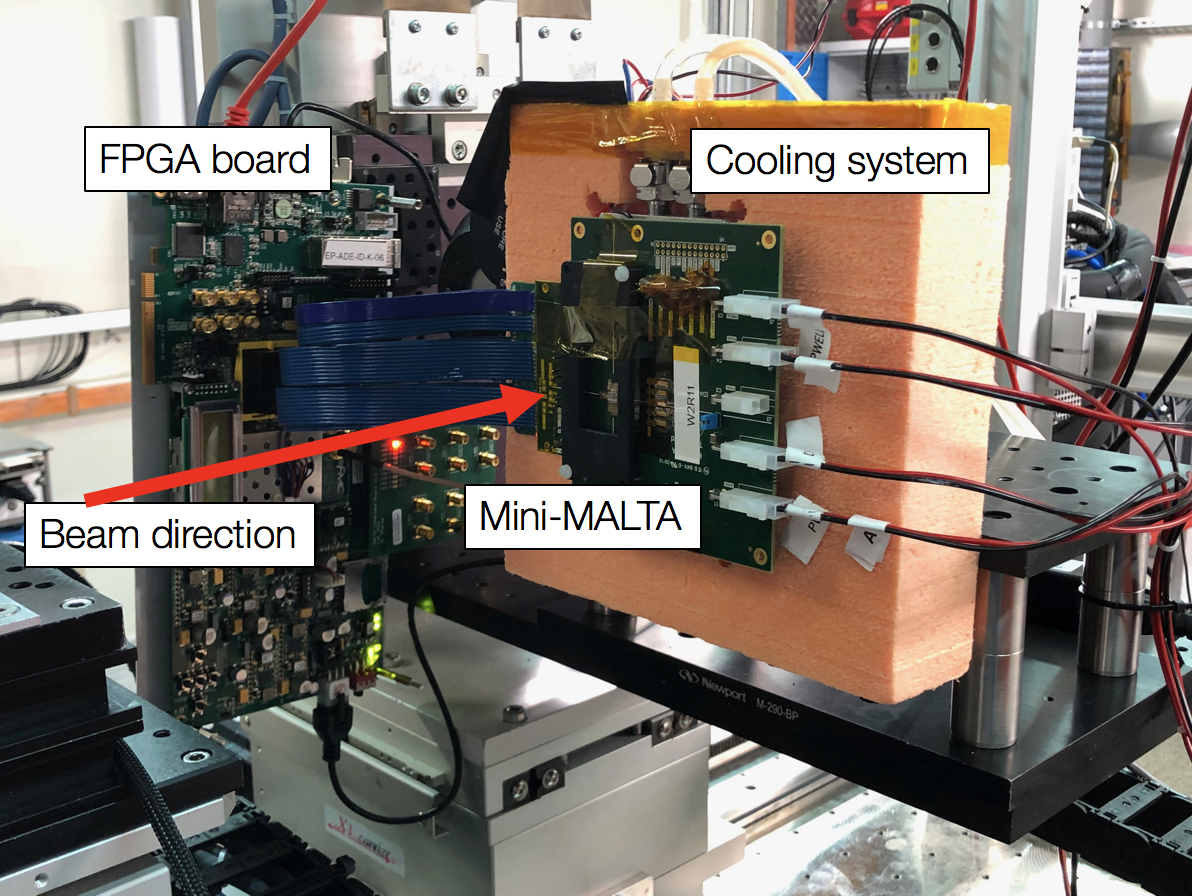}\label{fig:setup}}\qquad
\subfigure[CAD drawing of the cooling system with the attached mini-MALTA PCB.]{\includegraphics[width=0.35\textwidth]{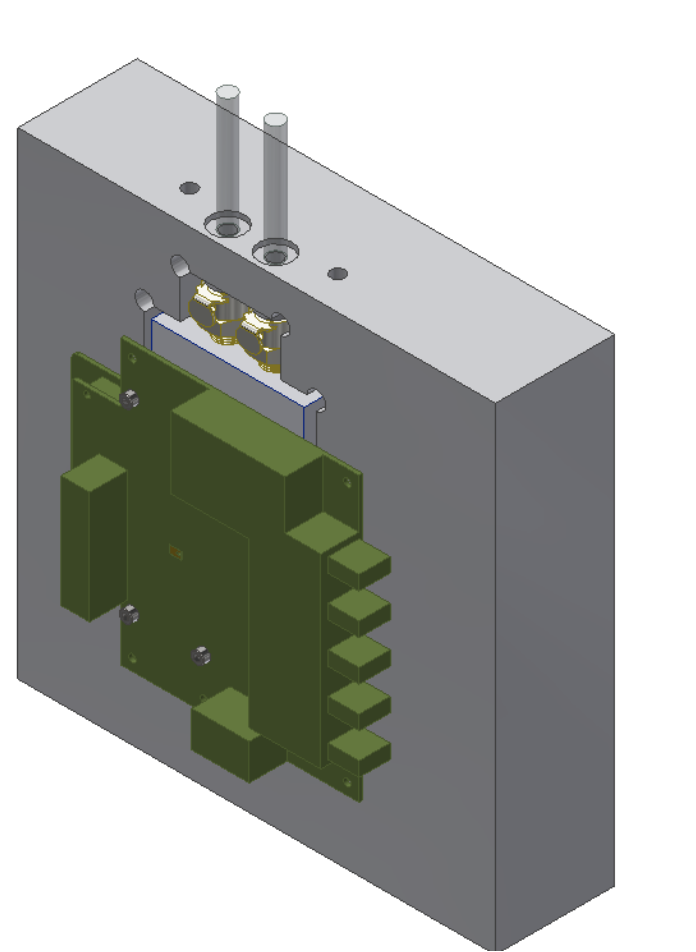}\label{fig:cooling}}
\caption{Experimental Setup in the beamline and design of the cooling system.}
\label{fig:beamline}
\end{figure}

The measurements were performed at the B16 beamline at Diamond Light Source. The setup in the beamline is shown in Figure \ref{fig:setup}. The mini-MALTA chip was mounted on a PCB, which was attached onto the cooling system. The cooling system was developed in Oxford and is shown in Figure \ref{fig:cooling}. It consisted of a water-cooled metal chuck, which reached a temperature of $\mathrm{10\;\degree C}$, coupled to the warm side of a Peltier element, which in turn was connected to a cold adapter plate onto which the mini-MALTA PCB was attached. This cooling system consistently kept the sensor at $\mathrm{-13\;\degree C}$, with the temperature being monitored with a temperature sensor.

The X-ray beam was focused using a Compound Reflective Lens (CRL) X-ray mirror arrangement to a beam spot of 2 $\mathrm{\mu m}$ (FWHM), as measured using a knife edge technique. The mini-MALTA chip was mounted on a motion stage, which was placed at the location of the focal point of the X-ray beam. The device was read out using a Kintex KC705 FPGA board. 

The choice of beam energy was made based on an estimate of how much energy a minimum ionising particle (MIP) would deposit in the depletion region of mini-MALTA. The stopping power for a MIP in silicon is \cite{PDG}:

\begin{equation}
  \label{eq:mip}
  \Big\langle \frac{dE}{dx}\Big\rangle=3.88\;\mathrm{\frac{MeV}{cm}}.
\end{equation}


Mini-MALTA is expected to have a depletion depth of between 20 and 25 $\mathrm{\mu m}$, depending on chip properties, biasing voltages and levels of irradiation. This means that a MIP would deposit between 7.7 and 9.7 keV in the device. X-rays interact primarily through the photoelectric effect and deposit all of their energy in one location. Based on this, a beam energy of 8 keV was chosen. 

The beam was attenuated by 0.5 mm of aluminium, in order to decrease the hit rate to that which can easily be read out by the mini-MALTA. During the measurements, the detector was moved in both orthogonal directions to the beam in 2 $\mathrm{\mu m}$ steps with a precision of 400 nm. Data was acquired for 1 s at each position to reach high statistics. The scans covered a total area of  100x100 $\mathrm{\mu m^2}$, which fully contained at least four pixels.

The measurements on the different devices were done at different thresholds, which are summarised in Table \ref{table:thresholds}. The thresholds were chosen based on the amount of noise in the chip, set with the discriminator bias current ("IDB") and measured with a threshold scan on each device. The value of the threshold does not influence the results of an X-ray testbeam as much as it does for a MIP testbeam, because the photon always deposits all of its energy when it interacts. For an 8 keV photon one would expect 2200 $\mathrm{e^-}$ to be produced, which is significantly higher than the thresholds noted in table \ref{table:thresholds}, even if the charge is split between multiple pixels. 

\begin{table}
\centering
\begin{tabularx}{\textwidth}{|X|X|X|X|X|X|}
 \hline
 Sample & Fluence $\mathrm{n_{eq}/cm^2}$ & TID (MRad)& MALTA threshold (e) & extra deep p-well threshold (e) &  $\mathrm{n^-} $ gap  threshold (e)\\ [0.5ex] 
 \hline\hline
 W2R11 & 0 & 0 &  368  & 197 & 190 \\ [1ex] \hline
 W2R9 & 5e14 (p) & 66 &  533 & 303 & 274\\[1ex]  \hline
 W2R1 & 1e15 (n) &  & 330 & 181 & 187\\[1ex]  \hline
 W5R9 & 5e14 (p) & 70 & 681 & 379 & 374\\ [1ex] \hline
 W4R9 & 7e13 (p) & 9 & 437 & 234 & 232\\[1ex]  \hline
 \hline
\end{tabularx}
\caption{Measured thresholds for the different samples and sectors.}
\label{table:thresholds}
\end{table}

\section{Results}


\subsection{Single Pixel Response \& Analysis}

\begin{figure}
\centering
\subfigure[Illustration of technique used for the analysis for a single pixel of the continuous $\mathrm{n^-}$  sector on W2R11. The green square outlines the centre of the pixel, which the number of hits is normalized to. The red square outlines the expected pixel area.]{\includegraphics[width=0.45\textwidth]{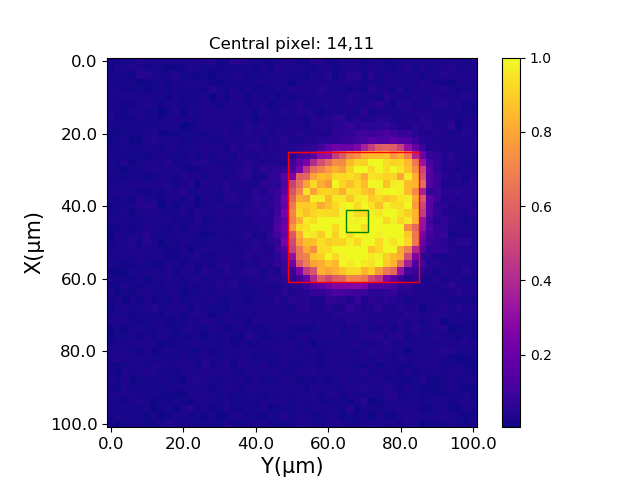}\label{fig:principle}}\qquad
\subfigure[Example of a summed pixel response map and profile of the response along the blue line for the extra deep p-well sector of W2R1. The response for each pixel is shown in a different color and the total normalised response is shown as a dashed line.]{  \includegraphics[width=0.45\textwidth]{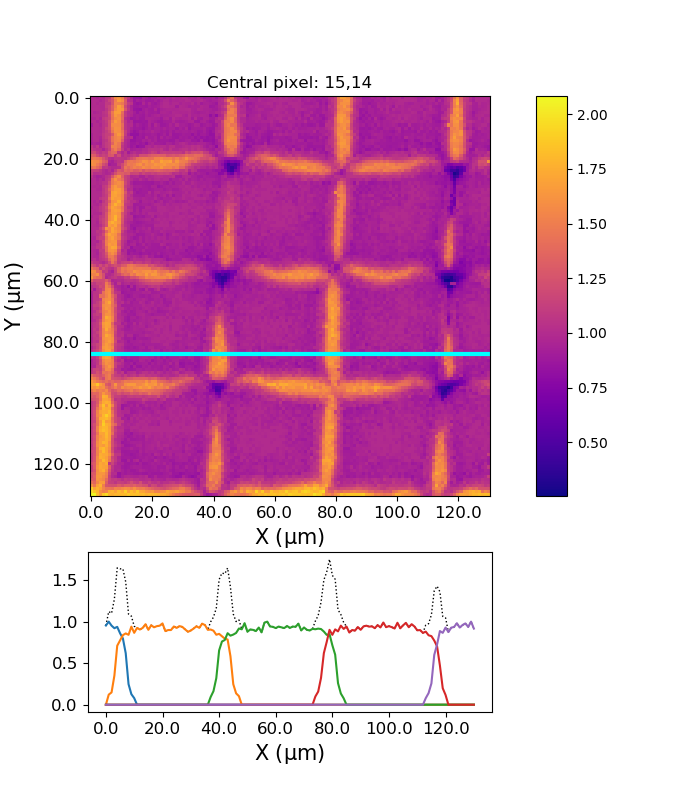}\label{fig:fullmap}}
\caption{Normalised response map for one pixel and for all pixels in one scan.}
\label{fig:analysis}
\end{figure}

The number of hits in a particular pixel was considered at each stage position. This was then stored in a 2D map, an example of which is shown in Figure \ref{fig:principle}. The pixel shape is clearly visible. Next, the centre of each pixel was identified by applying a Gaussian filter to the image and finding the location with the maximum value. Then, the average number of hits was calculated in a small region around the pixel centre, which is outlined with a green square. The number of hits in each pixel was then normalised to that average. This removes pixel to pixel variations. Taking the average of a larger region around the pixel centre also accounts for variations in the number of hits per step of the motion stage. Subsequently, the average normalised number of hits was calculated within a 36.4 x 36.4 $\mathrm{\mu m^2}$ square, which is shown in red in Figure \ref{fig:principle} and corresponds to the theoretical size of the pixel. This average is defined as the relative photon pixel response. The uncertainty on the pixel response is defined as the standard deviation of the number of hits in the normalisation area, i.e. the green square.

The calculation was repeated for each pixel in the scan and the pixel responses are averaged across the pixels which are fully visible in the scan. The single pixel maps are then added together, while removing everything from the pixel maps which is below 7\% normalised response, as those hits are likely to stem from the X-ray halo or noise in the detector. The pixel response calculations are done before this cut, so the normalisation and calculations are not affected by its choice, rather it is implemented to make sure the that the summed pixel response maps only show hits from the beam and not the X-ray halo. An example of the summed pixel response map is shown in Figure \ref{fig:fullmap}. At the pixel edges charge can be picked up by two pixels at the same time, which leads to a response above 1. This charge sharing is discussed in more detail in section \ref{subsec:chargesharing}. 

\begin{figure}
\centering
  \centering
  \includegraphics[width=0.5\textwidth]{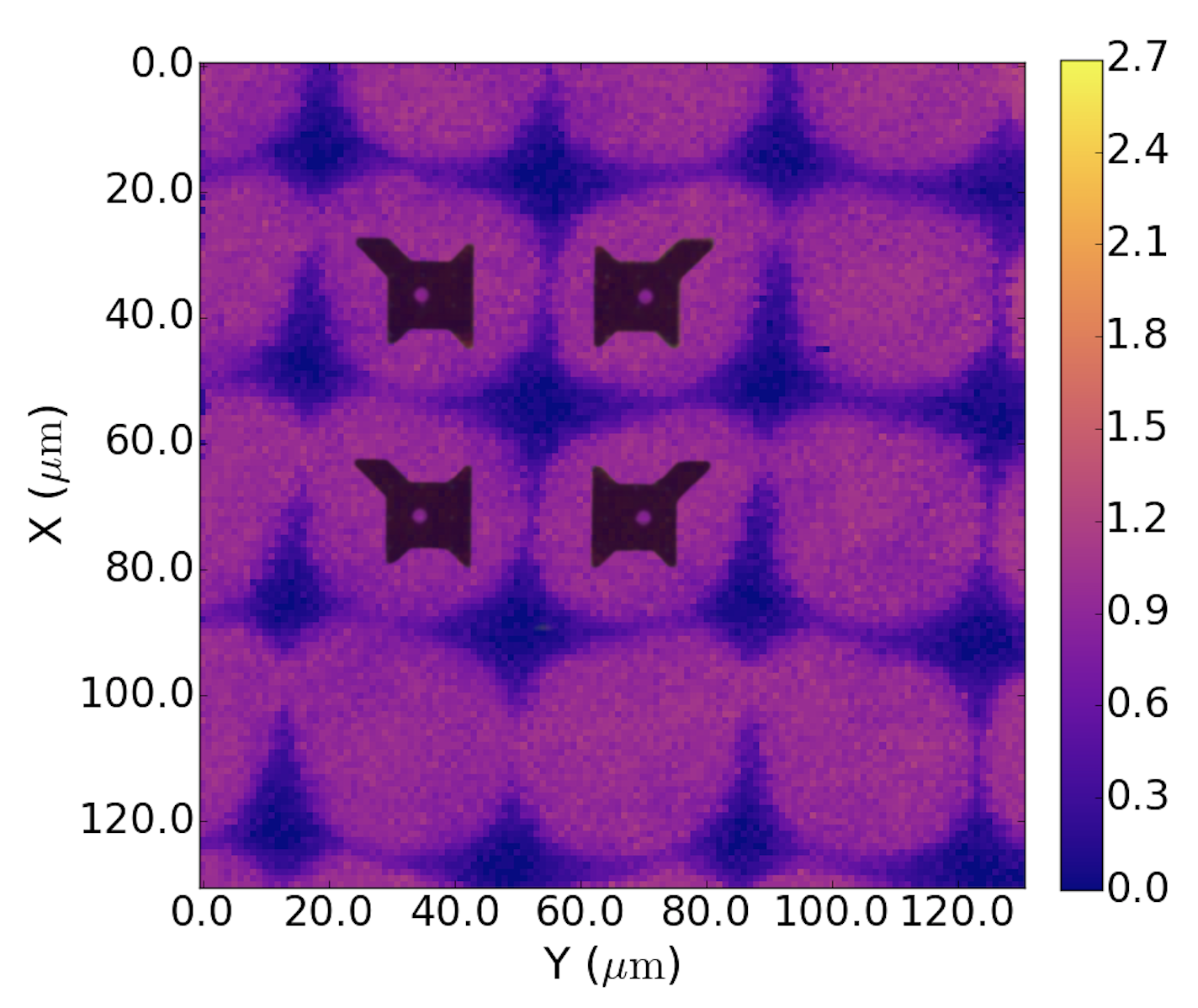}
  \caption{Overlay of the scan of the continuous $\mathrm{n^-}$  sector in W2R1 (irradiated to 1e15 $\mathrm{n_{eq}/cm^2}$) and the shape of the deep p-well cutout and collection electrodes. The cutout of the deep p-well influences the charge collection shape.}
  \label{fig:overlay}
\end{figure}

An interesting feature of the mini-MALTA is that the pixel response is asymmetric and extended in one direction. There is also a double column structure visible, i.e. two adjacent columns of pixels appear mirrored. This is in agreement with what was observed in previous testbeams of the MALTA chip. The reason for the asymmetric pixel shape is the shape of the deep p-well cut-out. This is the region between collection electrode and the p-well which does not contain p-type silicon. The shape of the deep p-well cut-out is shown as an overlay in Figure \ref{fig:overlay}, with the n-type collection electrode as the pink dot in the centre. In regions with less p-well coverage, i.e. a larger gap, one can observe a higher pixel response. This is caused by the fact that a larger gap causes a larger potential difference with respect to the collection electrode. The layouts of the p-well are mirrored in double-columns, which explains the double-column structure of the pixels.

\subsection{Pixel response as a function of dose}

\begin{figure}
\centering     
\subfigure[W2R11 continuous $\mathrm{n^-}$, unirradiated]{\label{fig:W2R11_MALTA}\includegraphics[width=0.32\textwidth]{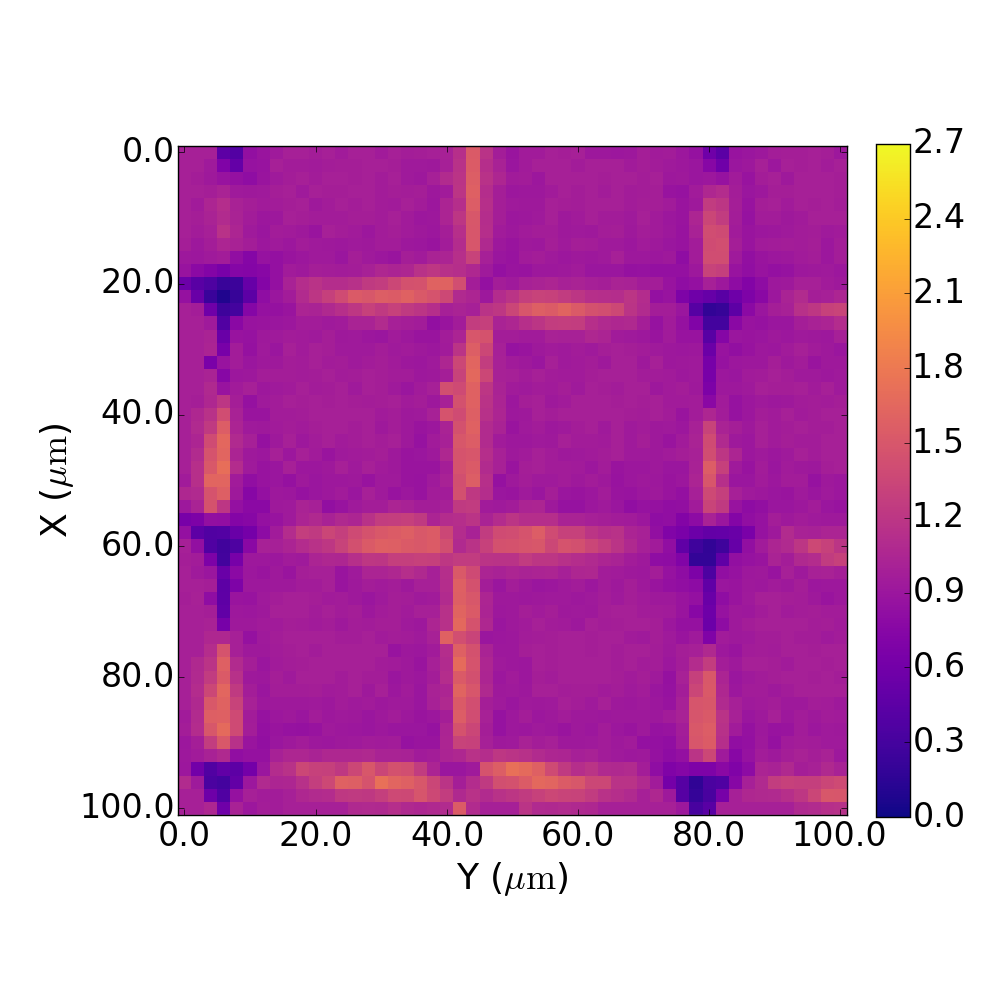}}
\subfigure[W2R11 extra deep p-well, unirradiated.]{\label{fig:W2R11_pwell}\includegraphics[width=0.32\textwidth]{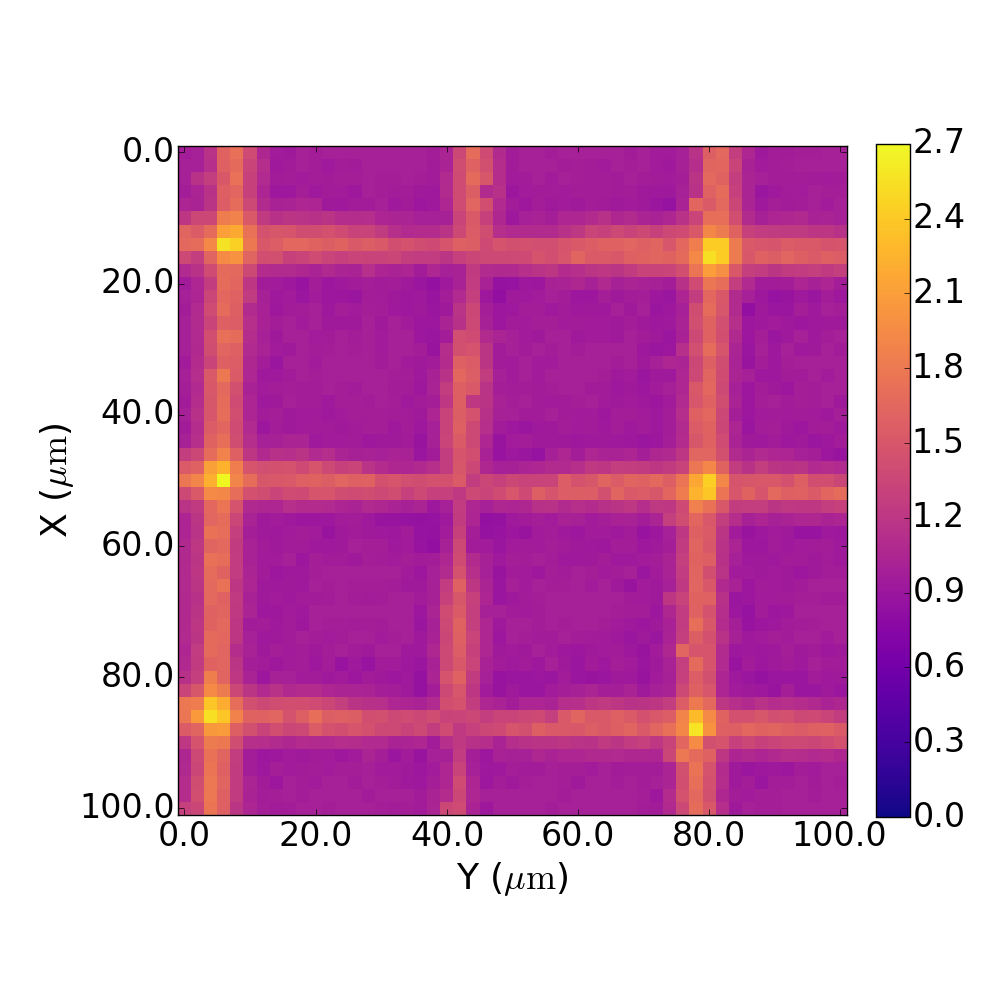}}
\subfigure[W2R11 $\mathrm{n^-}$ gap, unirradiated.]{\label{fig:W2R11_ngap}\includegraphics[width=0.32\textwidth]{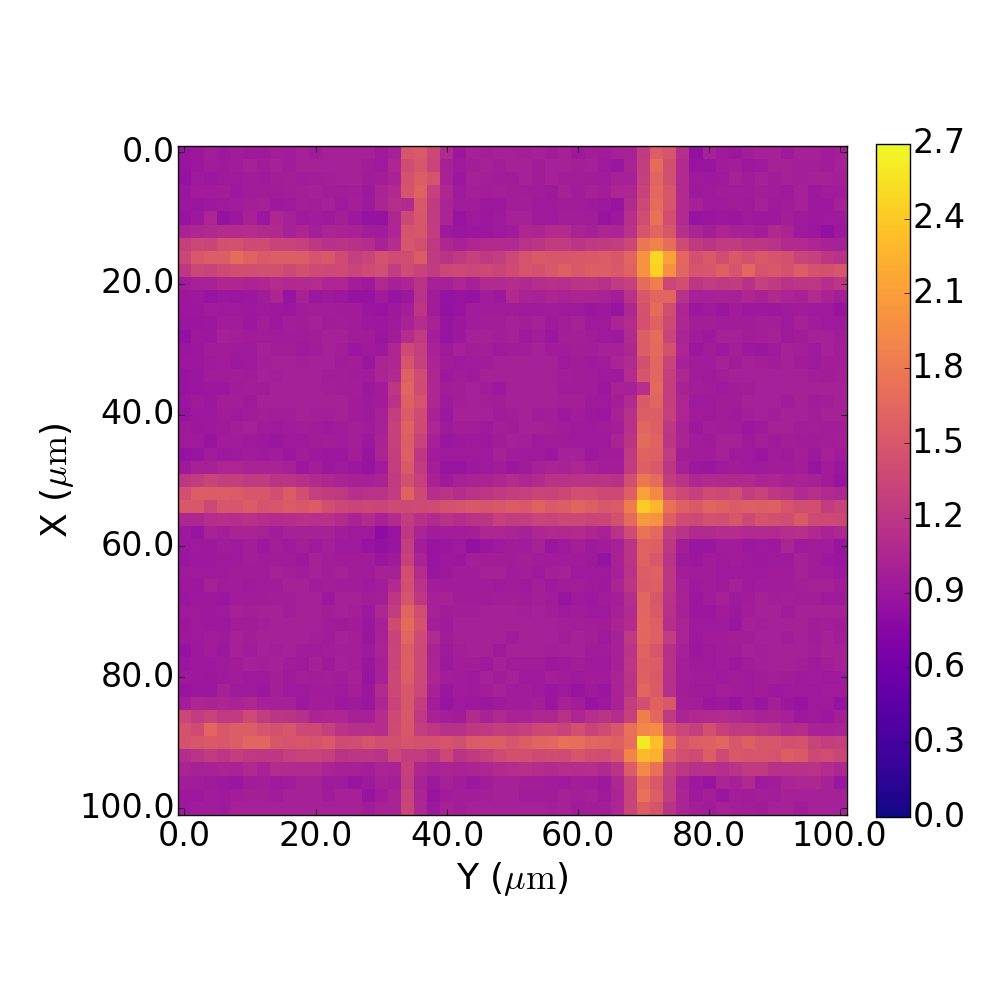}}
\subfigure[W2R9 continuous $\mathrm{n^-}$, 5e14 $\mathrm{n_{eq}/cm^2}$.]{\label{fig:W2R9_MALTA}\includegraphics[width=0.32\textwidth]{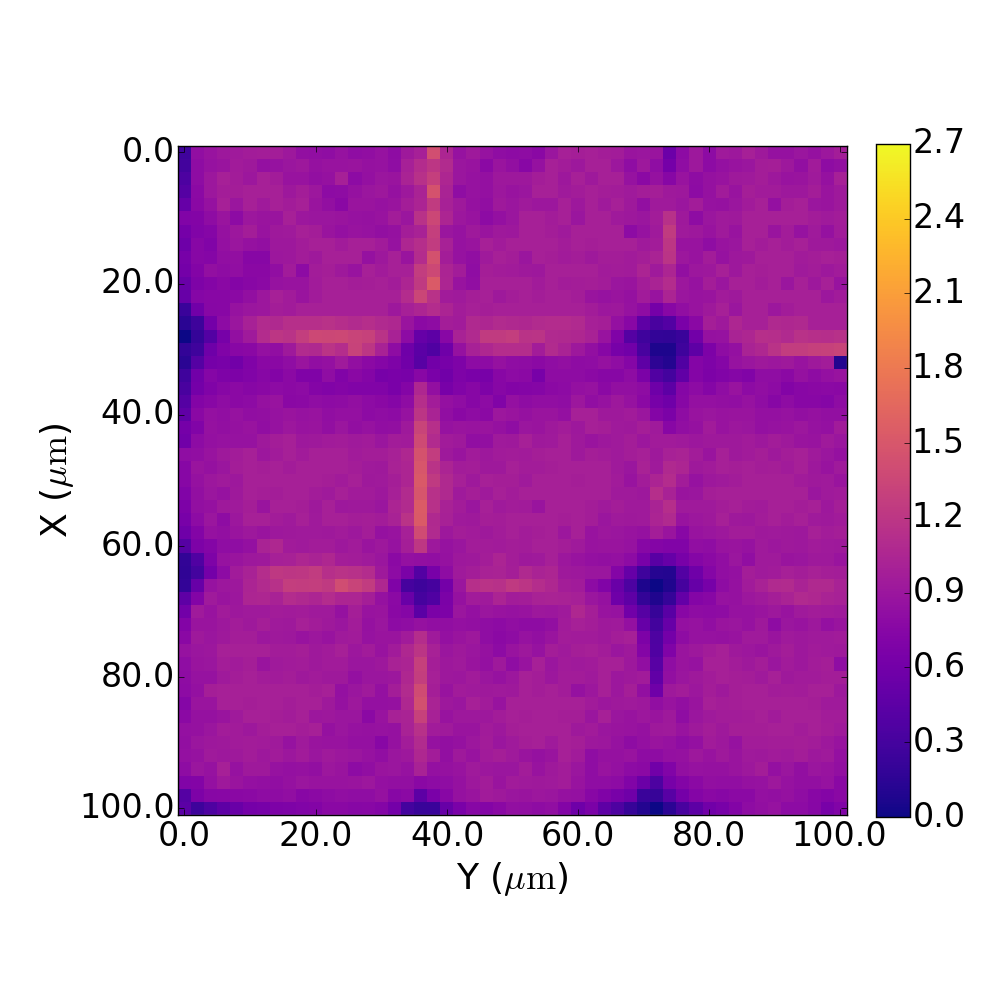}}
\subfigure[W2R9 extra deep p-well, 5e14 $\mathrm{n_{eq}/cm^2}$.]{\label{fig:W2R9_pwell}\includegraphics[width=0.32\textwidth]{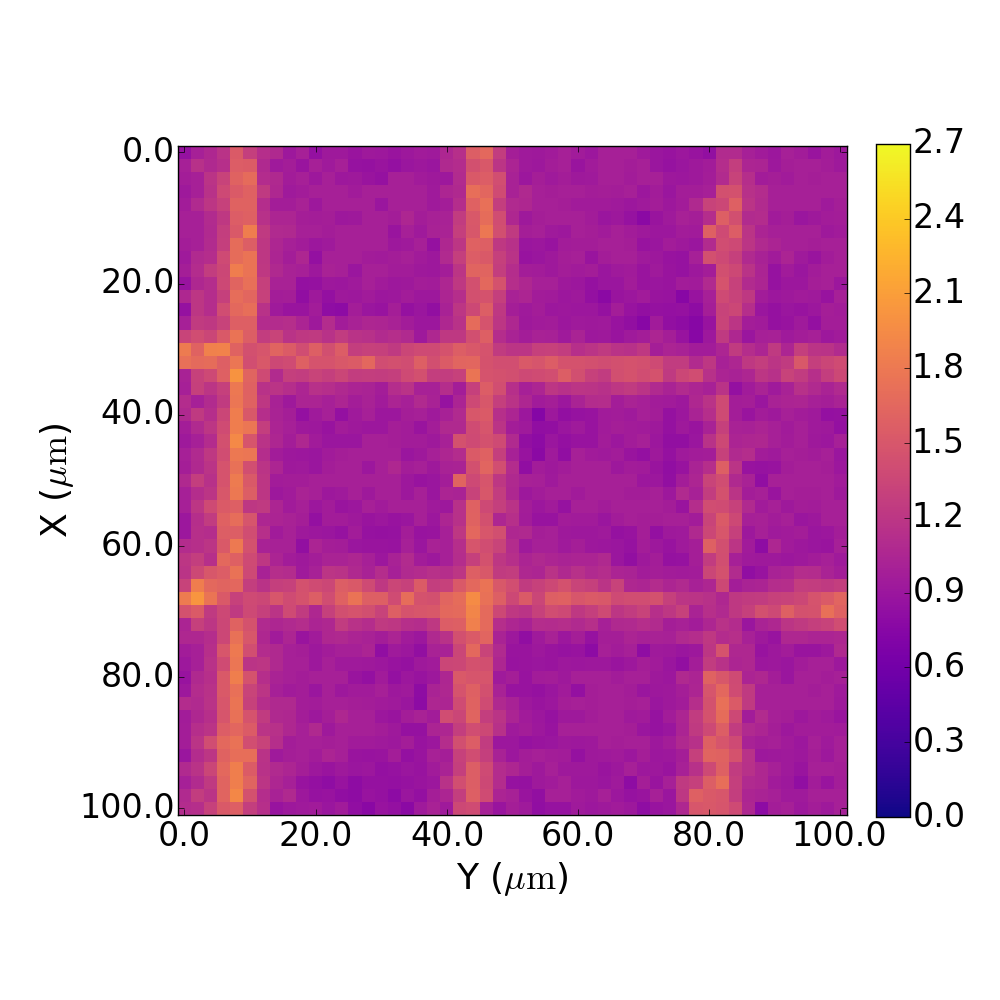}}
\subfigure[W2R9 $\mathrm{n^-} $ gap, 5e14 $\mathrm{n_{eq}/cm^2}$.]{\label{fig:W2R9_ngap}\includegraphics[width=0.32\textwidth]{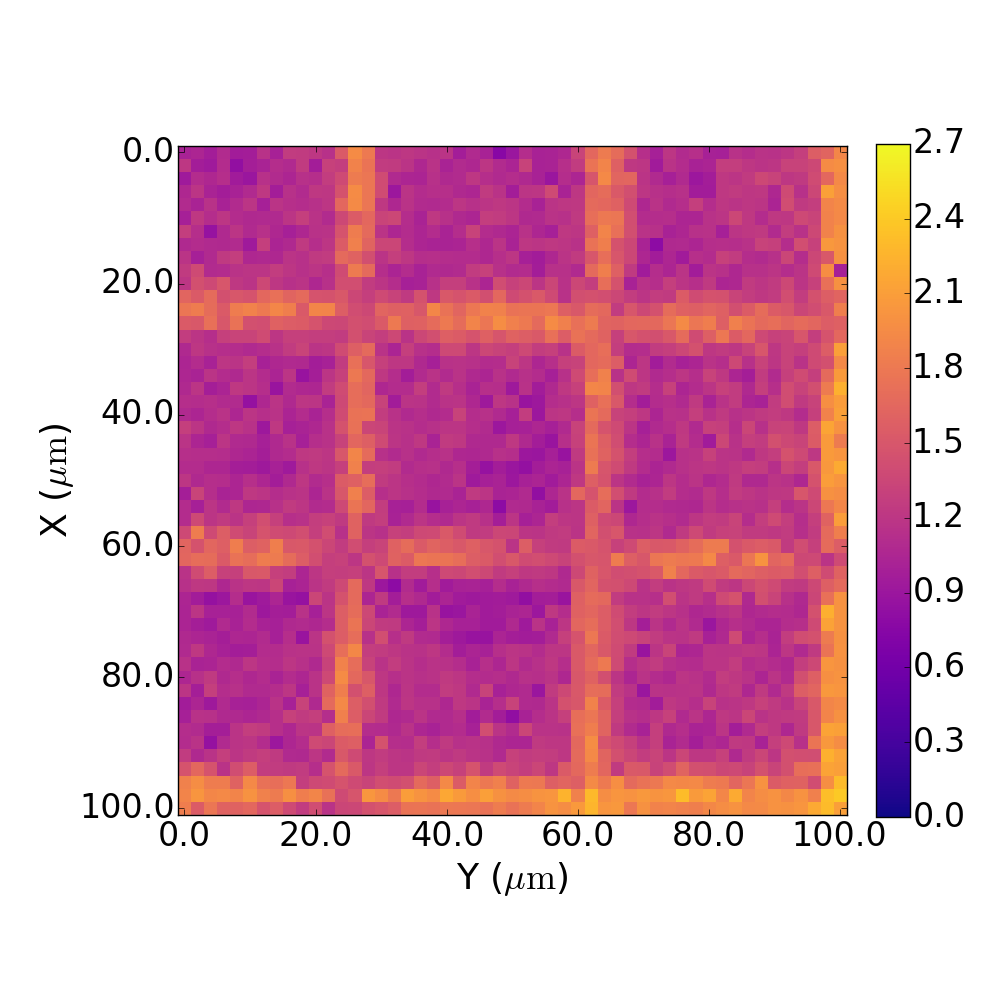}}
\subfigure[W2R1 continuous $\mathrm{n^-}$, 1e15 $\mathrm{n_{eq}/cm^2}$.]{\label{fig:W2R1_MALTA}\includegraphics[width=0.32\textwidth]{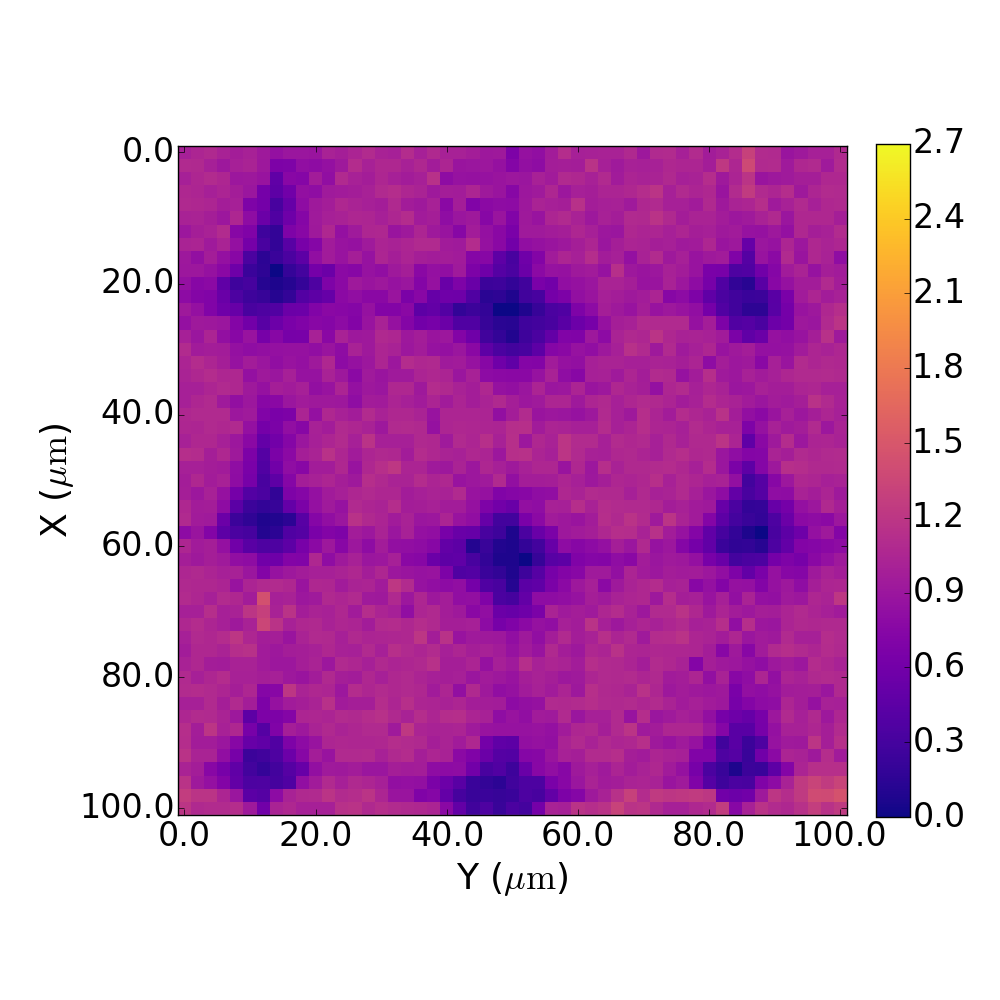}}
\subfigure[W2R1 extra deep p-well, 1e15 $\mathrm{n_{eq}/cm^2}$.]{\label{fig:W2R1_pwell}\includegraphics[width=0.32\textwidth]{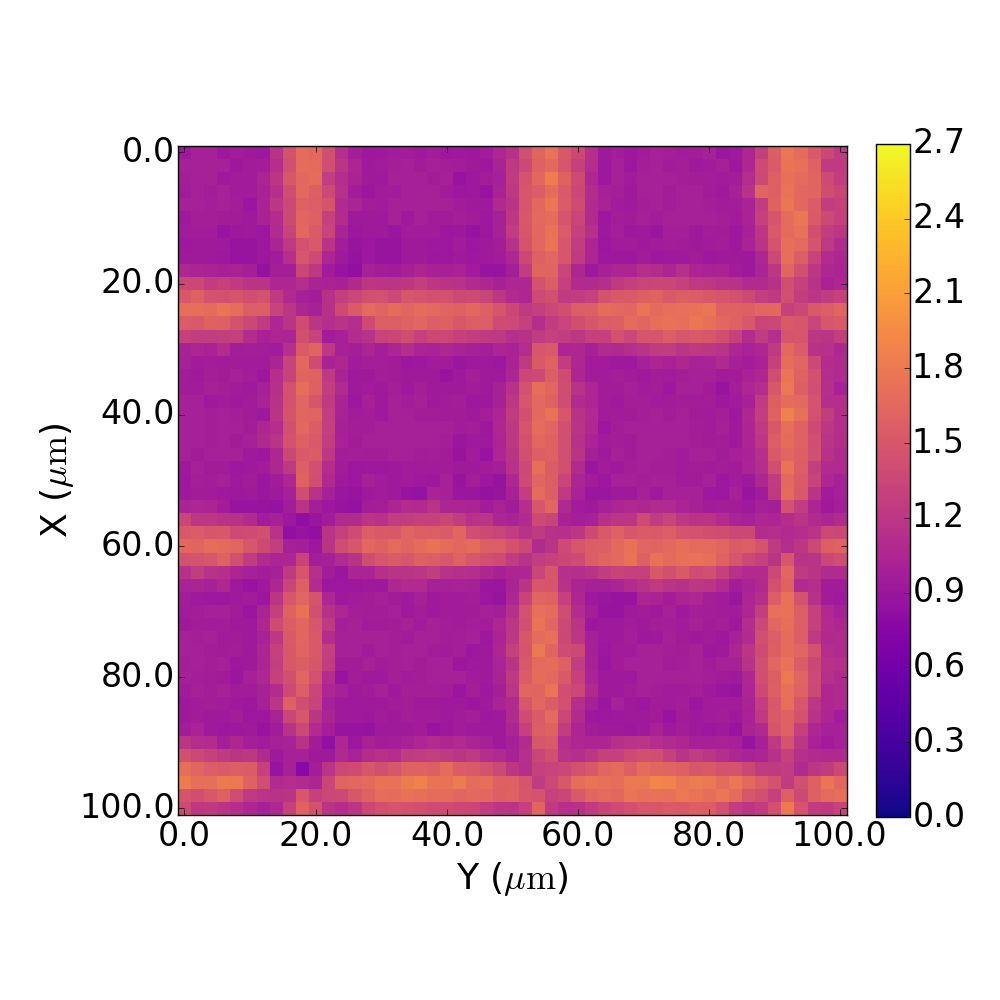}}
\subfigure[W2R1 $\mathrm{n^-} $ gap, 1e15 $\mathrm{n_{eq}/cm^2}$.]{\label{fig:W2R1_ngap}\includegraphics[width=0.32\textwidth]{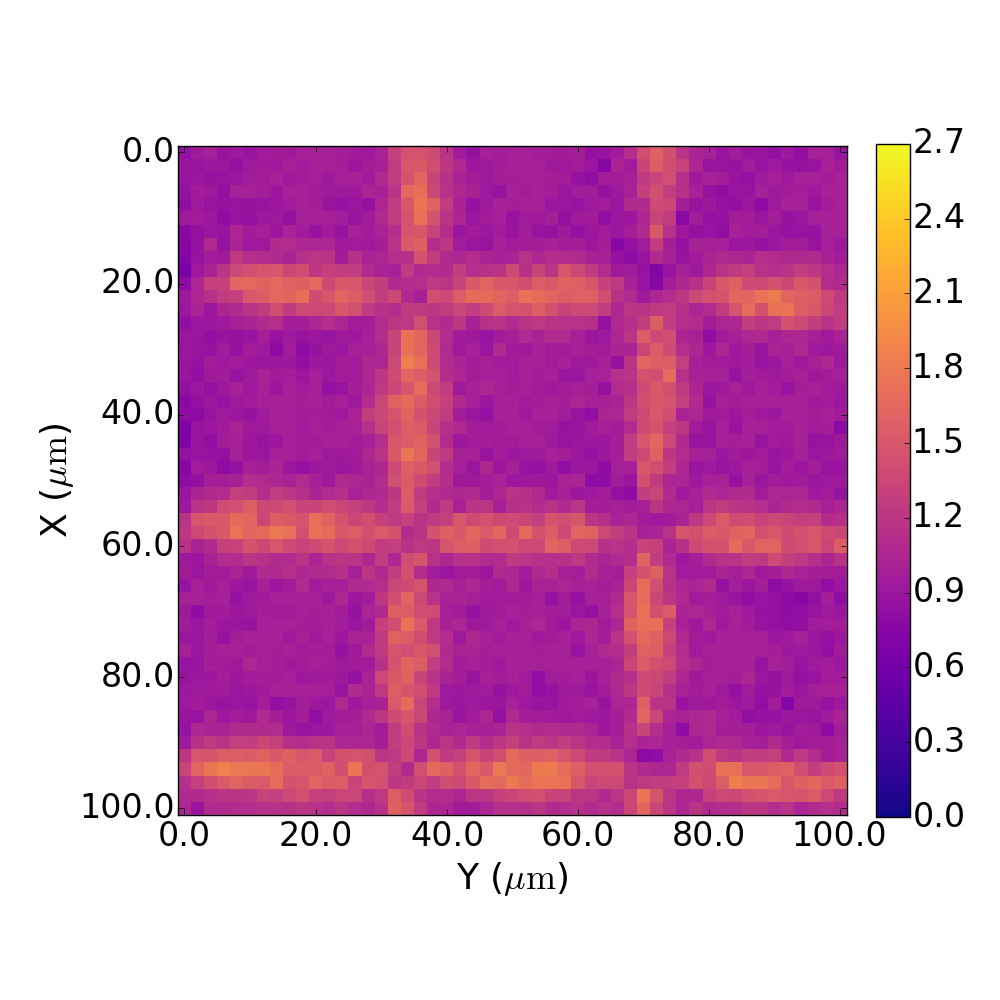}}
\caption{Pixel response maps for the different samples. For each sample the continuous $\mathrm{n^-}$, extra deep p-well and $\mathrm{n^-} $ gap sector are shown. W2R11 is unirradiated, W2R9 is proton-irradiated to 5e14  $\mathrm{n_{eq}/cm^2}$ and 66 Mrad and W2R1 is neutron-irradiated to 1e15 $\mathrm{n_{eq}/cm^2}$.}
\label{fig:EffResults}
\end{figure}

The results of the normalised pixel response analysis for the different samples are shown in Table \ref{table:efficiencies} and the response maps are shown in Figures \ref{fig:W2R11_MALTA} - \ref{fig:W2R1_ngap}. In order to extract a dependence on levels of irradiation, the measurements were performed at the same applied biasing voltage. The most accurate comparison between different designs and radiation levels is provided by comparing the samples W2R11, W2R9, and W2R1. These samples are from the same wafer (W2), which means they must have the same resistivity. 

The continuous $\mathrm{n^-}$ sectors for the unirradiated (W2R11) and neutron irradiated samples (W2R1) are shown in Figures \ref{fig:W2R11_MALTA} and \ref{fig:W2R1_MALTA} respectively. For the unirradiated sample, almost all of the pixel is fully responsive, except for some of the corners. For the neutron irradiated sample the pixel response decreases significantly around the corners and pixel edges. Overall the average pixel response in the continuous $\mathrm{n^-}$ sector decreases with neutron equivalent radiation dose, by more than 10\% between W2R11 and W2R1. The photon response which is found here is not directly comparable to the efficiency which is determined at proton testbeams, because of differences in the testbeam method and the different energy deposition mechanism for protons and photons. However, the proton efficiency determined at previous a MALTA testbeam at SPS \cite{MALTA_OLD} shows a similar trend with irradiation. The same behaviour was also observed a the mini-MALTA testbeam performed at ELSA \cite{MiniMALTA2019}. 

Compared to the continuous $\mathrm{n^-}$ design, the extra deep p-well and $\mathrm{n^-} $ gap designs perform better. For the unirradiated sample there is only a small improvement numerically, but in the pixel response maps (Figures \ref{fig:W2R11_pwell} and \ref{fig:W2R11_ngap}) there is no more loss in response at the pixel edges. Moreover, for the irradiated samples, there is almost no decrease in relative response observed with radiation dose. Most of the pixel remains fully responsive, which means the new designs significantly improve the detector performance after irradiation.

For the other samples (W5R9 and W4R9) the new designs show a similar improvement of the pixel response, compared to the MALTA design.

\begin{table}
\centering
\begin{tabularx}{\textwidth}{|X|X|X|X|X|X|}
 \hline
 Sample & Fluence $\mathrm{n_{eq}/cm^2}$ & TID (MRad)& continuous $\mathrm{n^-}$ response (\%) & extra deep p-well response (\%) & $\mathrm{n^-} $ gap response (\%)\\ [0.5ex] 
 \hline\hline
 W2R11 & 0 & 0 &  $88.3\pm2.4$  & $90.5\pm2.2$ & $90.9\pm2.2$ \\ [1ex] \hline
 W2R9 & 5e14 (p) & 66 &  $81.2\pm2.8$ & $87.6\pm4.2$ & $88.4\pm3.8$\\[1ex]  \hline
 W2R1 & 1e15 (n) &  & $75.4\pm3.8$ & $90.5\pm2.8$ & $89.0\pm3.1$\\[1ex]  \hline
 W5R9 & 5e14 (p) & 70 & $80.4\pm2.8$ & $89.0\pm2.5$ & $89.3\pm2.2$\\ [1ex] \hline
 W4R9 & 7e13 (p) & 9 & $78.7\pm2.6$ & $89.8\pm2.3$ & $89.9\pm2.3$\\[1ex]  \hline
 \hline
\end{tabularx}
\caption{Calculated average pixel responses for the different samples and designs.}
\label{table:efficiencies}
\end{table}

\subsection{Pixel response as a function of biasing voltage}

\begin{figure}
\centering
\subfigure[Continuous $\mathrm{n^-}$ sector in W2R1 at -6 V biasing voltage.]{\includegraphics[width=0.45\textwidth]{figures/006050.png}\label{fig:6V}}\qquad
\subfigure[Continuous $\mathrm{n^-}$ sector in W2R1 at -20 V biasing voltage.]{  \includegraphics[width=0.45\textwidth]{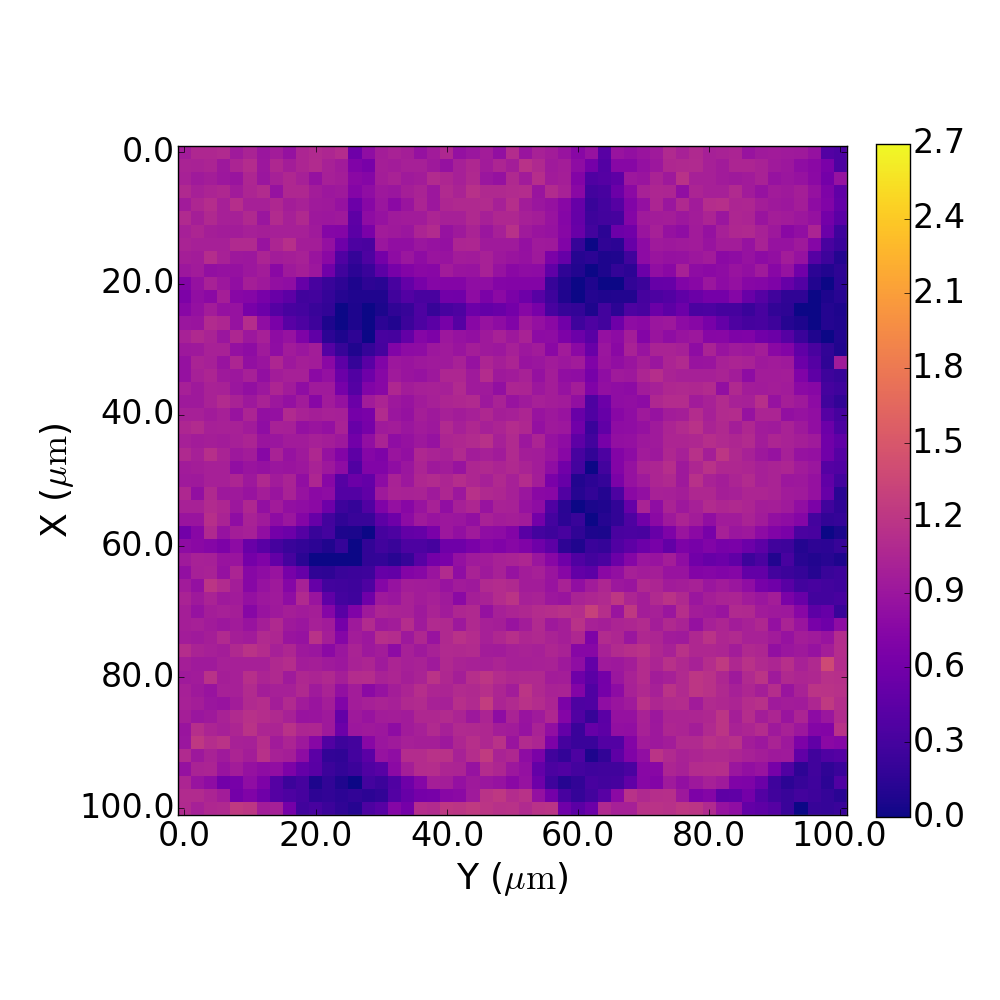}\label{fig:20V}}
\caption{Pixel response maps for continuous $\mathrm{n^-}$ sector in the W2R1 sample (neutron irradiated to 1e15 $\mathrm{n_{eq}/cm^2}$) at -6 and -20 V biasing voltage.}
\label{fig:biasing}
\end{figure}

\begin{figure}
\centering
\subfigure[$\mathrm{n^-}$ gap sector in W2R1 at -6 V biasing voltage.]{\includegraphics[width=0.45\textwidth]{figures/006049.png}\label{fig:Ngap6V}}\qquad
\subfigure[$\mathrm{n^-}$ gap sector in W2R1 at -20 V biasing voltage.]{  \includegraphics[width=0.45\textwidth]{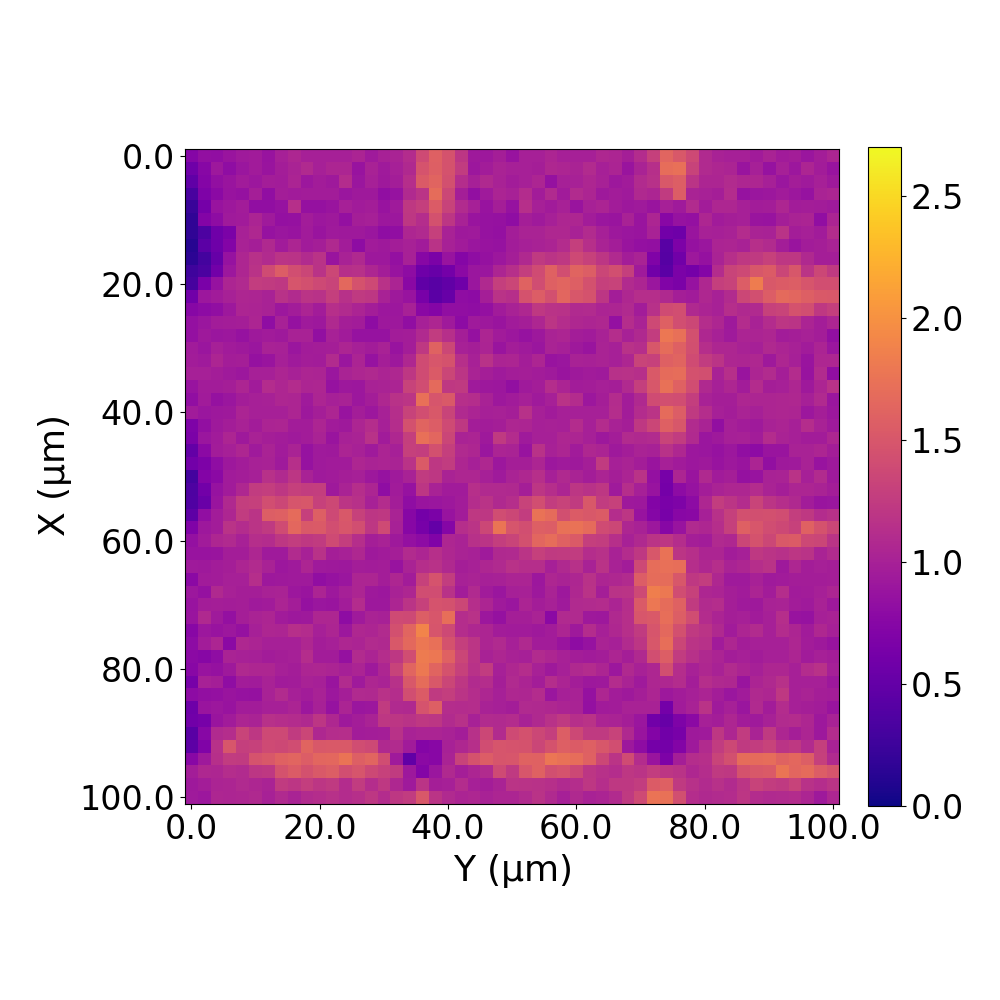}\label{fig:Ngap20V}}
\caption{Pixel response maps for $\mathrm{n^-}$ gap sector in the W2R1 sample (neutron irradiated to 1e15 $\mathrm{n_{eq}/cm^2}$) at -6 and -20 V biasing voltage.}
\label{fig:biasingNgap}
\end{figure}

An additional measurement was performed to compare the pixel response at different biasing voltages. Each of the sectors was scanned once at -6 V and -20 V. The resultant response maps for the continuous $\mathrm{n^-}$ sector at each voltage are shown in Figure \ref{fig:20V}. The average pixel response decreases from 76.7\% to 72.2\% with increasing bias voltage, which means that the pixel edges become less responsive with respect to the pixel centre and appear sharper. The same effect is seen for the new designs, as shown for the $\mathrm{n^-}$ gap design in Figure \ref{fig:biasingNgap}, and the response results summarised in Table \ref{table:bias20}. Simulations suggest that there are two effects caused by a higher applied voltage: a longer drift path and faster drift along the sensor depth \cite{TCAD}. In the pixel center the faster drift causes better charge collection. At the pixel edges however, the charge deposited below the deep p-well implant is pushed towards the deep p-well quickly and has to travel almost parallel to the surface. This longer drift path leads to a higher probability for the charge to be trapped and not to reach the collection electrode. And as the drift velocity perpendicular to the surface does not increase with higher voltage, more charge is lost and the normalised response around the pixel edges decreases. 


\begin{table}
\centering
\begin{tabularx}{\textwidth}{|X|X|X|X|X|X|}
 \hline
 Sample & Bias (V) & continuous $\mathrm{n^-}$ \mbox{response (\%)} & extra deep p-well \mbox{response (\%)} & $\mathrm{n^-}$ gap \mbox{response (\%)}\\ [0.5ex] 
 \hline\hline
 W2R1 & -6 & $76.7\pm3.8$ & $91.1\pm3.0$ & $90.0\pm3.1$\\[1ex]  \hline
 & -20 & $72.2\pm3.3$ & $86.6\pm3.9$ & $86.4\pm2.9$\\[1ex]  \hline
 \hline
\end{tabularx}
\caption{Average normalised response for the different sectors of W2R1 (neutron irradiated to 1e15 $\mathrm{n_{eq}/cm^2}$) for different biasing voltages.}
\label{table:bias20}
\end{table}



\subsection{Charge Sharing}
\label{subsec:chargesharing}

 Charge sharing occurs when a particle hit occurs in the region between two pixels and the charge is collected and registered as a signal by both of those pixels. To quantify the amount of charge sharing, the pixel response outside of the nominal pixel area was considered. First, the sum of the normalised number of hits outside of the pixel area, i.e. the red square in Figure \ref{fig:principle}, was calculated. Then the total normalized number of hits was found for each pixel and the ratio of the two was defined as the charge sharing percentage.  Additionally, the extent of the charge sharing region was found. This was defined as the distance from the nominal pixel edge where one could still see hits in the pixel. The average of that distance was found for each pixel edge and averaged among all fully visible pixels. The uncertainty on the charge sharing extents is defined as the standard deviation of the averaged values, and as the pixel shape is asymmetric it is thus relatively large.

The results for the charge sharing analysis are shown in Tables \ref{table:chargepercentage} and \ref{table:chargeextent}, summarising the charge sharing percentages and extents respectively. For the percentages no error on the individual values was calculated, but due to the observed variation in hits from step to step, a systematic uncertainty of around 2\% is expected. Additionally, the difference between the thresholds in different sectors introduces a 0.3\% uncertainty on the response calculated at the pixel edges, and thus a small systematic uncertainty on the charge sharing. Both results are consistent with each other, i.e. when the percentage decreases so does the average extent of the region. For the samples from the same wafer, the charge sharing decreases with increased irradiation in the continuous $\mathrm{n^-}$ sector, which is in agreement with what can be seen on the pixel response maps, i.e. the response loss around the pixel edges. For the extra deep p-well and $\mathrm{n^-}$ gap sector the charge sharing does not decrease and a there might even be a marginal increase in charge sharing extents after irradiation.


\begin{table}
\centering
\begin{tabularx}{\textwidth}{|X|X|X|X|X|X|}
 \hline
 Sample & Fluence $\mathrm{n_{eq}/cm^2}$ & TID (Mrad)  & continuous $\mathrm{n^-}$ charge \mbox{sharing (\%)} & p-well charge \mbox{sharing (\%)} & $\mathrm{n^-}$ gap charge \mbox{sharing (\%)}\\ [0.5ex] 
 \hline\hline
 W2R11 & 0 & 0 & $15.8$  & $20.3$ & $16.9$ \\ [1ex] \hline
 W2R9 & 5e14 (p) & 66 & $7.7$ & $17.9$ & $21.2$\\[1ex]  \hline
 W2R1 & 1e15 (n) &  & $7.1$ & $23.6$ & $21.3$\\[1ex]  \hline
 W5R9 & 5e14 (p) & 70 & $7.3$ & $14.1$ & $15.4$\\ [1ex] \hline
 W4R9 & 7e13 (p) & 9 & $8.4$ & $14.9$ & $15.7$\\[1ex]  \hline
 \hline
\end{tabularx}
\caption{Calculated charge sharing percentages for the different samples and designs.}
\label{table:chargepercentage}
\end{table}

\begin{table}
\centering
\begin{tabularx}{\textwidth}{|X|X|X|X|X|X|}
 \hline
 Sample & Fluence $\mathrm{n_{eq}/cm^2}$ & TID (Mrad) &  continuous $\mathrm{n^-}$ charge sharing ($\mathrm{\mu m}$) & extra deep p-well charge sharing ($\mathrm{\mu m}$) & $\mathrm{n^-}$ gap charge sharing ($\mathrm{\mu m}$)\\ [0.5ex] 
 \hline\hline
 W2R11 & 0 & 0 & $4.5\pm1.5$  & $5.4\pm1.3$ & $4.5\pm1.3$ \\ [1ex] \hline
 W2R9 & 5e14 (p) & 66 &  $2.6\pm1.8$ & $5.4\pm2.2$ & $5.5\pm2.0$\\[1ex]  \hline
 W2R1 & 1e15 (n) & &  $3.2\pm2.0$ & $6.1\pm2.0$ & $6.5\pm2.6$\\[1ex]  \hline
 W5R9 & 5e14 (p) & 70 & $2.4\pm2.0$ & $4.5\pm1.8$ & $4.5\pm2.7$\\ [1ex] \hline
 W4R9 & 7e13 (p) & 9 &  $2.4\pm1.9$ & $4.1\pm1.9$ & $4.1\pm1.0$\\[1ex]  \hline
 \hline
\end{tabularx}
\caption{Extents of the charge sharing regions for the different samples and designs.}
\label{table:chargeextent}
\end{table}

As the final step of the charge sharing analysis, the asymmetry of the charge sharing was quantified. The asymmetric pixel shape and double column structure of mini-MALTA lead to an asymmetry in the charge sharing. In particular this can be seen in Figures \ref{fig:W2R11_pwell} and \ref{fig:W2R11_ngap}, where only every second column has charge sharing regions with values above 2. 

\begin{figure}
\centering
  \centering
  \includegraphics[width=0.5\textwidth]{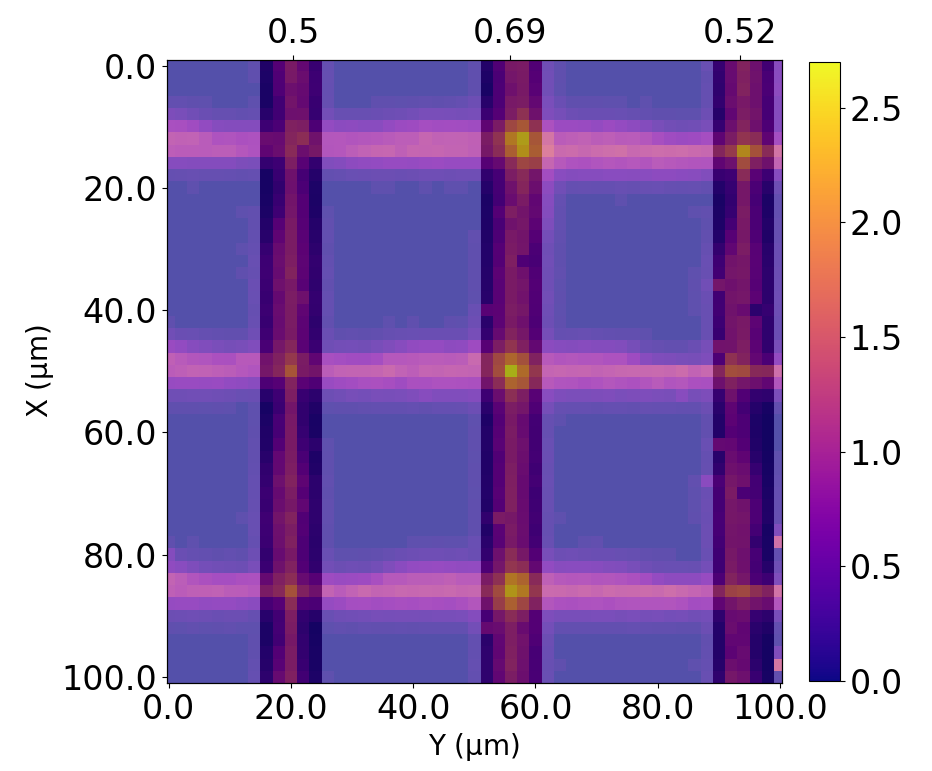}
  \caption{Illustration of the charge sharing asymmetry analysis. The response due to charge sharing is calculated in the grey columns and their ratio is defined as the charge sharing asymmetry.}
  \label{fig:asymmetry}
\end{figure}

To quantify this asymmetry, the normalised pixel response outside of the nominal pixel area was considered. This is shown as a map in Figure \ref{fig:asymmetry}. The average response due to charge sharing was calculated within 10 $\mathrm{\mu m}$ wide columns around the pixel edges, shown in grey. Clearly the central bin in Figure \ref{fig:asymmetry} has a higher response. The asymmetry of the charge sharing is then defined as the ratio of the response of the double-columns with these higher values and those with lower values. The error on the asymmetry is calculated from the variation of the response in the different columns of the same type (i.e. high or low charge sharing) and the normalisation error from the previous pixel response calculation. The results for the charge sharing asymmetry are shown in Table \ref{table:chargeasymmetry}. The asymmetry is smaller for the extra deep p-well and $\mathrm{n^-}$ gap designs compared to the continuous $\mathrm{n^-}$ layer one. There is a decrease of charge sharing asymmetry with irradiation, which could be explained by the radiation damage which causes the charge sharing regions to become broader, thus being more important than the asymmetric pixel shape.

\begin{table}
\centering
\begin{tabularx}{\textwidth}{|X|X|X|X|X|X|}
 \hline
 Sample & Fluence $\mathrm{n_{eq}/cm^2}$ & TID (Mrad) &  continuous $\mathrm{n^-}$ \mbox{asymmetry} & extra deep p-well \mbox{asymmetry} & $\mathrm{n^-}$ gap \mbox{asymmetry}\\ [0.5ex] 
 \hline\hline
 W2R11 & 0 & 0 & $1.80\pm0.11$  & $1.49\pm0.07$ & $1.43\pm0.07$ \\ [1ex] \hline
 W2R9 & 5e14 (p) & 66 &  $1.49\pm0.12$ & $1.11\pm0.10$ & $1.17\pm0.09$\\[1ex]  \hline
 W2R1 & 1e15 (n) & &  $1.16\pm0.10$ & $1.07\pm0.06$ & $1.15\pm0.09$\\[1ex]  \hline
 W5R9 & 5e14 (p) & 70 & $1.39\pm0.09$ & $1.02\pm0.06$ & $1.12\pm0.07$\\ [1ex] \hline
 W4R9 & 7e13 (p) & 9 &  $1.22\pm0.08$ & $1.02\pm0.05$ & $1.09\pm0.07$\\[1ex]  \hline
 \hline
\end{tabularx}
\caption{Charge sharing asymmetries for the different samples and designs.}
\label{table:chargeasymmetry}
\end{table}

\subsection{Clustering Analysis}

The photon response maps show the superposition of two different effects. They firstly contain information about the shape and depth of the depletion region and the charge collection, which explain the response loss in the corners of the continuous $\mathrm{n^-}$ sector. Secondly, they provide information about the cluster sizes in the chip, i.e. how many pixels see a photon deposited at a particular location. This is then related to the charge sharing, which was discussed section \ref{subsec:chargesharing}. To separate these two effects a cluster analysis was performed on the data. 

\begin{figure}
\centering     
\subfigure[W2R11 continuous $\mathrm{n^-}$, unirradiated]{\label{fig:W2R11_MALTA_cluster}\includegraphics[width=0.32\textwidth]{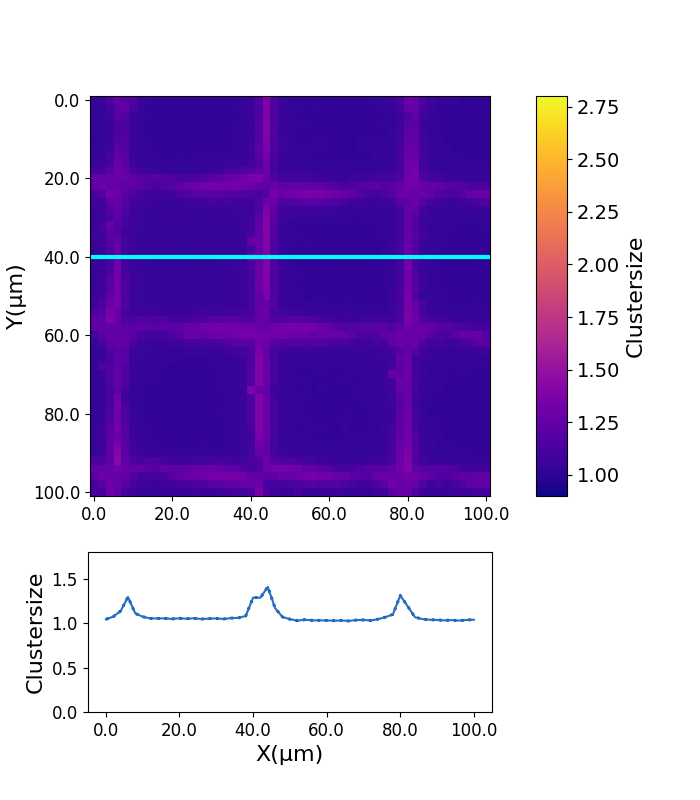}}
\subfigure[W2R11 extra deep p-well, unirradiated.]{\label{fig:W2R11_pwell_cluster}\includegraphics[width=0.32\textwidth]{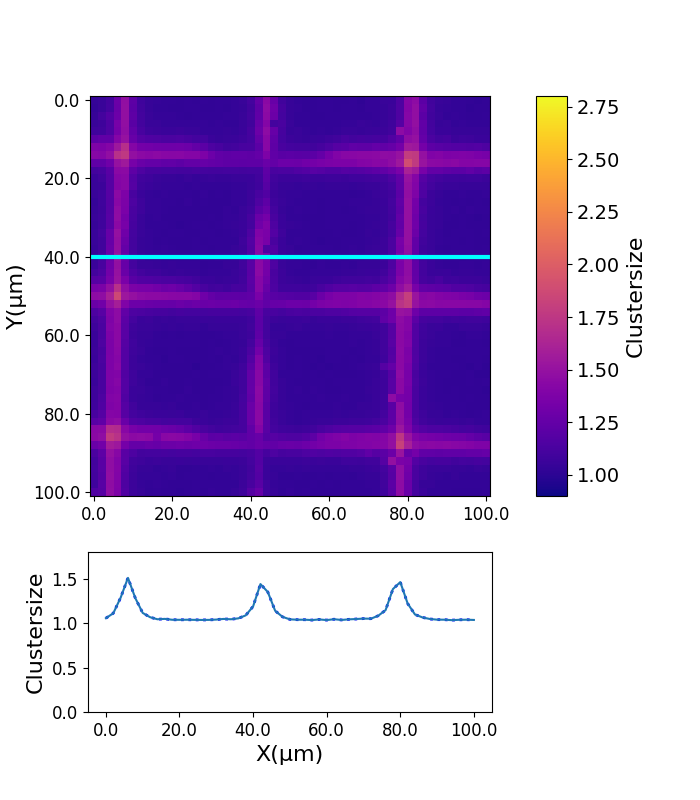}}
\subfigure[W2R11 $\mathrm{n^-}$ gap, unirradiated.]{\label{fig:W2R11_ngap_cluster}\includegraphics[width=0.32\textwidth]{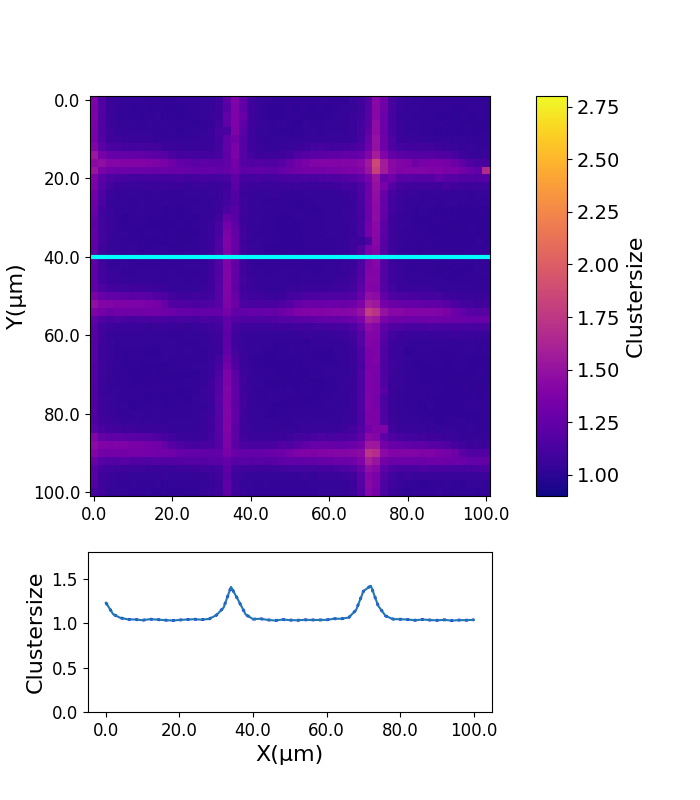}}
\subfigure[W2R1 continuous $\mathrm{n^-}$, 1e15 $\mathrm{n_{eq}/cm^2}$.]{\label{fig:W2R1_MALTA_cluster}\includegraphics[width=0.32\textwidth]{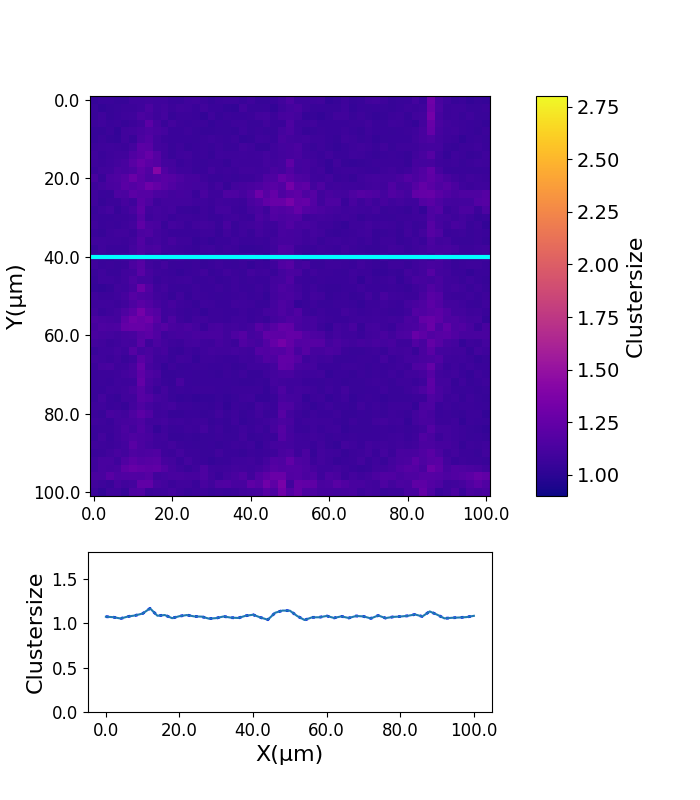}}
\subfigure[W2R1 extra deep p-well, 1e15 $\mathrm{n_{eq}/cm^2}$.]{\label{fig:W2R1_pwell_cluster}\includegraphics[width=0.32\textwidth]{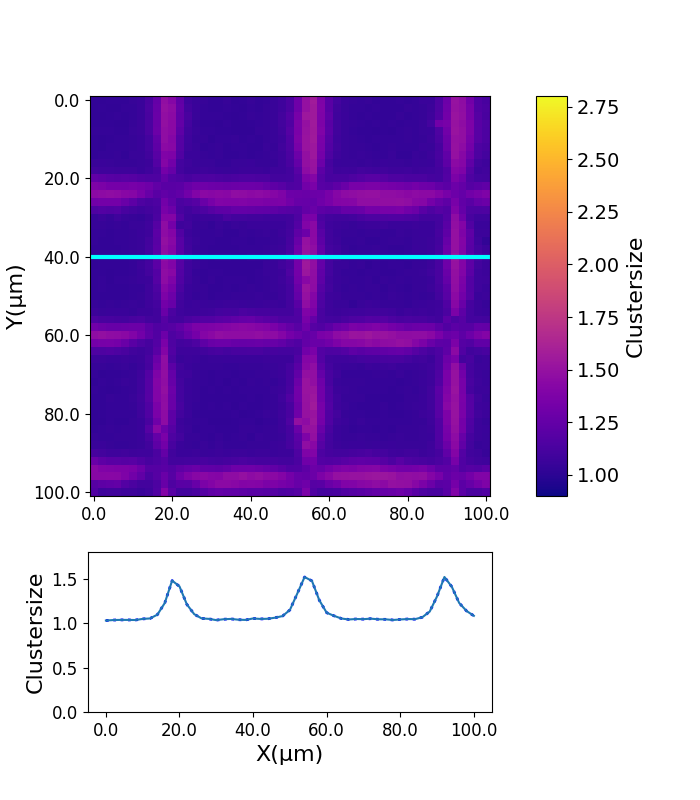}}
\subfigure[W2R1 $\mathrm{n^-}$ gap, 1e15 $\mathrm{n_{eq}/cm^2}$.]{\label{fig:W2R1_ngap_cluster}\includegraphics[width=0.32\textwidth]{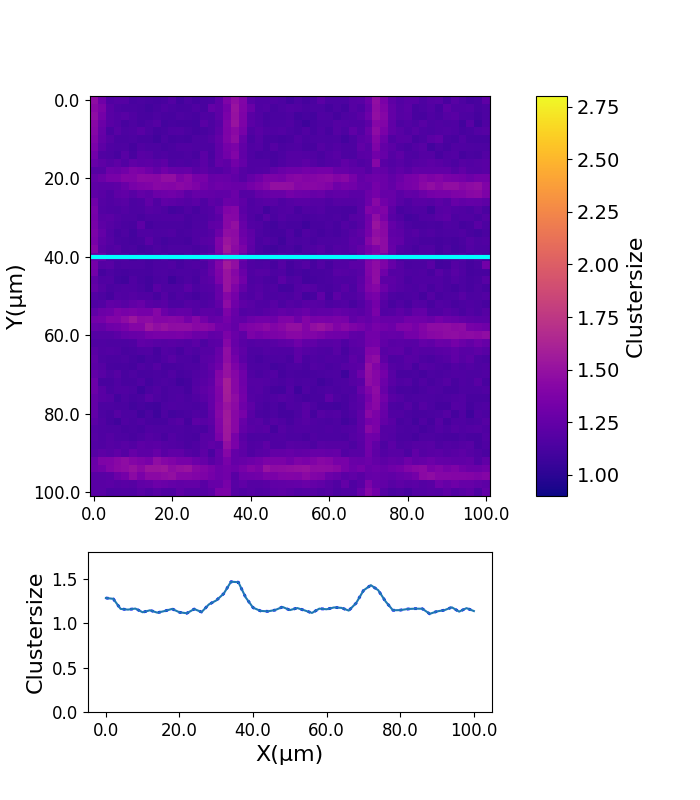}}
\caption{Map of cluster sizes for the different sectors in the W2R11 (unirradiated) and W2R1 (neutron-irradiated to 1e15 $\mathrm{n_{eq}/cm^2}$) sample. The bottom panel shows the profile of cluster size for a particular x location, which is indicated by a blue line.}
\label{fig:ClusterResults}
\end{figure}

The mini-MALTA chip records events in 25 ns windows. In the clustering analysis each of these events was considered individually and the cluster size was found, i.e. the number of neighbouring pixels which show hits. Then for each stage position the average cluster size was calculated.
This was only done for all of the visible pixels in a scan, to reduce hits from the X-ray halo. There was also a cut applied on the total number of hits at each stage position: only the pixels with at least 1\% of the maximum number of hits were considered in the calculation of the average cluster size. The resulting average cluster size was plotted as a function of stage position, with the results shown in Figure \ref{fig:ClusterResults}. A two-dimensional map is shown, as well as a profile of the cluster size at a particular x position.

As expected, the cluster size is 1 in the center of each pixel and then increases at the edges and corners. For the extra deep p-well and $\mathrm{n^-}$ gap sectors the resulting clustering maps look very similar to the pixel response maps, as the latter are dominated by charge sharing for these designs. The increase in charge sharing extents with irradiation can also be seen in the clustering maps. For the continuous $\mathrm{n^-}$ sector, average cluster sizes above 1 are also observed at the pixel edges, though the values are lower compared to the extra deep p-well and $\mathrm{n^-}$ gap sector. This effect is prominent in the photon response maps, as these are dominated by the loss in pixel response at the pixel edges due to depletion depth and charge collection. This analysis shows that there is also charge sharing present in the continuous $\mathrm{n^-}$ sector. The width of the charge sharing regions also increase with irradiation, but the average cluster size decreases.

\section{Conclusions}

The mini-MALTA prototype was tested using an 8 keV X-ray beam at Diamond Light Source. A beam spot with a size of 2 $\mathrm{\mu m}$ was scanned across the surface of the chip in 2 $\mathrm{\mu m}$ steps. From the number of hits in each pixel at each stage position it was possible to determine the in-pixel photon response. Devices with different levels of irradiation were compared and a decrease in pixel response with irradiation was observed for the standard continuous $\mathrm{n^-}$ layer design. The two new mini-MALTA designs with an extra deep p-well implant or a gap in the $\mathrm{n^-}$ layer performed better than the standard design and showed almost no decrease in pixel response with irradiation. The dependence of pixel response on substrate voltage was studied and a decrease in pixel response at high voltages was found.  The amount of charge sharing was quantified and found to be consistent with the response results and theoretical expectations.


\section{Acknowledgements}
This project has received funding from the European Union’s Horizon 2020 Research and Innovation programme under Grant Agreement no. 654168. (IJS, Ljubljana, Slovenia). This research project has been supported by the Marie Sklodowska-Curie Innovative Training Network of the European Commission Horizon 2020 Programme under contract number 675587 "STREAM".
Dr. Ben Phoenix, Prof. David Parker, Amelia Hunter, and the operators at the MC40 cyclotron in Birmingham (UK).
We acknowledge Diamond Light Source for time on Beamline B16 under Proposal MM2206-1.

\bibliographystyle{JHEP}

\bibliography{bibliography}

\end{document}